\newlength{\absize}
\documentstyle[12pt]{article}
\renewcommand{\baselinestretch}{1.5}
\renewcommand{\arraystretch}{1.5}
\begin{document}
\thispagestyle{empty}
\pagestyle{empty}
\renewcommand{\thefootnote}{\fnsymbol{footnote}}
\newcommand{\starttext}{\newpage\normalsize
\pagestyle{plain}
\setlength{\baselineskip}{4ex}\par
\setcounter{footnote}{0}
\renewcommand{\thefootnote}{\arabic{footnote}}
}

\newcommand{\figsize}{}

\input prepictex
\input pictex
\input postpictex
\newdimen\tdim
\tdim=\unitlength
\def\stpltsmbl{\setplotsymbol ({\small .})}
\def\tarrow{\arrow <5\tdim> [.3,.6]}

\newcommand{\preprint}[1]{\begin{flushright}
\setlength{\baselineskip}{3ex}#1\end{flushright}}
\renewcommand{\title}[1]{\begin{center}\LARGE
#1\end{center}\par}
\renewcommand{\author}[1]{\vspace{2ex}{\Large\begin{center}
\setlength{\baselineskip}{3ex}#1\par\end{center}}}
\renewcommand{\thanks}[1]{\footnote{#1}}
\renewcommand{\abstract}[1]{\vspace{2ex}\normalsize\begin{center}
\centerline{\bf Abstract}\par\vspace{2ex}\parbox{\absize}{#1
\setlength{\baselineskip}{2.5ex}\par}
\end{center}}

\newcommand{\segment}[2]{\put#1{\circle*{2}}}
\newcommand{\fig}[1]{figure~\ref{#1}} \newcommand{\hc}{{\rm h.c.}}
\newcommand{\ds}{\displaystyle} \newcommand{\eqr}[1]{(\ref{#1})}
\newcommand{\tr}{\,{\rm tr}} \newcommand{\uone}{{U(1)}}
\newcommand{\su}[1]{{SU(#1)}} \newcommand{\stu}{\su2\times\uone}
\newcommand{\be}{\begin{equation}} \newcommand{\ee}{\end{equation}}
\newcommand{\bp}{\begin{picture}} \newcommand{\ep}{\end{picture}}
\def\spur#1{\mathord{\not\mathrel{#1}}}
\def\lta{\mathrel{\displaystyle\mathop{\kern 0pt <}_{\raise .3ex
\hbox{$\sim$}}}} \def\gta{\mathrel{\displaystyle\mathop{\kern 0pt
>}_{\raise .3ex \hbox{$\sim$}}}} \newcommand{\sechead}[1]{\medskip{\bf
#1}\par\bigskip} \newcommand{\ba}[1]{\begin{array}{#1}\ds }
\newcommand{\cra}{\\ \ds} \newcommand{\ea}{\end{array}}
\newcommand{\forto}[3]{\;{\rm for}\; #1 = #2 \;{\rm to}\; #3}
\newcommand{\for}{\;{\rm for}\;} \newcommand{\cross }{\hbox{$\times$}}
\newcommand{\ol}{\overline} \newcommand{\bra}[1]{\left\langle #1 \right|}
\newcommand{\ket}[1]{\left| #1 \right\rangle}
\newcommand{\braket}[2]{\left\langle #1 \left|#2\right\rangle\right.}
\newcommand{\braketr}[2]{\left.\left\langle #1 right|#2\right\rangle}
\newcommand{\g}[1]{\gamma_{#1}} \newcommand{\half}{{1\over 2}}
\newcommand{\del}{\partial} \newcommand{\grad}{\vec\del}
\newcommand{\real}{{\rm Re\,}} \newcommand{\imag}{{\rm Im\,}}
\newcommand{\cl}[1]{\begin{center} #1\end{center}} \newcommand\etal{{\it et
al.}} \newcommand{\prl}[3]{Phys. Rev. Letters {\bf #1} (#2) #3}
\newcommand{\prd}[3]{Phys. Rev. {\bf D#1} (#2) #3}
\newcommand{\npb}[3]{Nucl. Phys. {\bf B#1} (#2) #3}
\newcommand{\plb}[3]{Phys. Lett. {\bf #1B} (#2) #3} \newcommand{\ie}{{\it
i.e.}} \newcommand{\etc}{{\it etc.\/}} \def\cA{{\cal A}} \def\cB{{\cal B}}
\def\cC{{\cal C}} \def\cD{{\cal D}} \def\cE{{\cal E}} \def\cF{{\cal F}}
\def\cG{{\cal G}} \def\cH{{\cal H}} \def\cI{{\cal I}} \def\cJ{{\cal J}}
\def\cK{{\cal K}} \def\cL{{\cal L}} \def\cM{{\cal M}} \def\cN{{\cal N}}
\def\cO{{\cal O}} \def\cP{{\cal P}} \def\cQ{{\cal Q}} \def\cR{{\cal R}}
\def\cS{{\cal S}} \def\cT{{\cal T}} \def\cU{{\cal U}} \def\cV{{\cal V}}
\def\cW{{\cal W}} \def\cX{{\cal X}} \def\cY{{\cal Y}} \def\cZ{{\cal Z}}
\renewcommand{\baselinestretch}{1.5} \renewcommand{\arraystretch}{1.5}
\newcommand{\boxit}[1]{\ba{|c|}\hline #1 \\ \hline\ea}
\newcommand{\mini}[1]{\begin{minipage}[t]{20em}{#1}\vspace{.5em}
\end{minipage}} \newcommand{\Q}{{_{\!Q}}} \newcommand{\U}{{_{\!U}}}
\newcommand{\D}{{_{\!D}}} \renewcommand{\L}{{_{\!L}}}
\newcommand{\X}{{_{\!X}}} \newcommand{\W}{{_{\!W}}}
\newcommand{\Z}{{_{\!Z}}}


\preprint{\#HUTP-96/A024\\ 6/96}
\title{
Decays of a Leptophobic Gauge Boson
\thanks{Research
supported in part by the National Science Foundation under Grant
\#PHY-9218167.}
}
\author{
Howard Georgi and Sheldon L. Glashow \\
Lyman Laboratory of Physics \\
Harvard University \\
Cambridge, MA 02138 \\
}
\date{}
\abstract{
We discuss the theory and phenomenology of decays of a leptophobic
$U(1)_\X$
gauge boson $X$, such as has been proposed to explain
the alleged deviations of
$R_b$ and $R_c$ from standard model predictions. 
If the scalars
involved in the breaking of the $SU(2)\times U(1)$ symmetry 
are sufficiently light, $X$ will sometimes
decay into a charged (or neutral) scalar along with an oppositely-charged
$W$ (or $Z$). These decay modes
could yield clean signals for
the leptophobic gauge bosons at hadron colliders and provide an
interesting window into the Higgs sector of the theory.
}
\starttext

\section{\label{intro} Introduction}

Several recent papers propose the existence of a
gauge boson $X$ that couples to quarks but not 
leptons~\cite{altarelli,chiappetta,babu,hinchliffe}. Such a boson 
could explain certain deviations from the standard model reported at LEP.
The $X$ boson, even if quite light, may have escaped detection
because its decay into quark-antiquark has a large QCD
background~\cite{cdf}. We focus on the model of reference \cite{hinchliffe}
and suggest the possible detection of the $X$ boson
via its decay into $W$ or $Z$ bosons plus scalars decaying
into heavy quarks. A search for these decay modes could provide evidence
for leptophobic gauge bosons produced at hadron colliders.

\section{\label{model} The Model}

The gauge group is the standard model $\stu$ supplemented by a $U(1)_X$
that does not act on the lepton fields. Left-handed quark doublets carry
$U(1)_\X$ quantum number $q_\Q$, right-handed $U$ quarks $q_\U$, and 
right-handed $D$ quarks $q_\D$. If the $SU(2)$ symmetry breaking 
is done by fundamental Higgs bosons then, in general, three Higgs doublets
are needed to generate quark and lepton masses. Their
Yukawa couplings have the form:
\be
\ol Q\,h_\U\,\tilde H_\U\, U
+\ol Q\,h_\D\, H_\D\, D
+\ol L\,h_\L\, H_{_L}\, E\;,
\label{keterm}
\ee
where the $h_j$ are Yukawa coupling matrices and $\tilde H=-
\sigma_2\,H^{\ds *}$. It follows that the $U(1)_X$ quantum numbers of the 
doublets (which have weak hypercharge is ${1\over2}$) are
\be
q(H_\U) = q_\U-q_\Q\,,
\quad\quad
q(H_\D) = q_\Q-q_\D\,,
\quad\quad
q(H_{_{\!L}}) = 0\,.
\ee
For special values of the $q_j$'s, it is possible to give mass to quarks
and leptons with only two Higgs doublets, but none of these special choices
were favored in the analysis of reference \cite{hinchliffe}. Thus we assume
that\footnote{Note that this rules out some otherwise interesting
possibilities, such as the $\eta$-model of \protect\cite{babu}.}
\be
0\neq q_\U-q_\Q\neq q_\Q-q_\D\neq0\,.
\ee
If electromagnetic gauge invariance is to be left unbroken, the Higgs
doublets
may be written in the form: 
\be
H_j=
\pmatrix{
\xi_j^+\cr
(\xi^0_j+i\,\pi^0_j+v_j)/\sqrt2\cr
}
\ee
for $j=U$, $D$, or $L$. With an appropriate choice of phases for the $H_j$
fields,
their VEVs
$v_j$ may be made real and positive.

For simplicity, and because this is what was assumed in
\cite{hinchliffe}, we assume that the symmetry breaking is done 
primarily by the VEVs
of the $U$ and $D$ doublets, and that the VEV of the $L$ doublet is
negligible, $v_\L\ll v_\U,v_\D$.
The extra $U(1)_\X$ couplings lead to gauge anomalies, so that additional
fermion states must be introduced to cancel them. This is discussed, for
example, in \cite{hinchliffe}. 

The relevant interaction arises via the Higgs mechanism from
the kinetic energy terms of the $H_\U$ and $H_\D$ doublets:
\be
D^\mu H_\U^\dagger\,D_\mu H_\U
+D^\mu H_\D^\dagger\,D_\mu H_\D
\label{ke}
\ee
where
\be
D^\mu=\partial^\mu+ig_2{\vec\tau\over2}\cdot\vec W^\mu
+ig_1{1\over2}\,B^\mu\mp ig_x(q_\Q-q_{_{U,D}})\,X^\mu
\label{covariant}
\ee
and $X^\mu$ is the new $U(1)_X$ gauge field.

To avoid large corrections to the standard model properties of the $Z$ we
must tune the parameters of the model to make the
$X$-$Z$ mixing small. This requires that
\be
g_{_{\!X}}^2\left|
v_\U^2(q_\U-q_\Q)
+v_\D^2(q_\Q-q_\D)
\right|\ll g_2^2v^2\;.
\label{smallmixing}
\ee
Of course, the primary motivation for models of this kind is that a small
amount of mixing can modify the $Z$ couplings slightly and result in a
better
fit to data than the unadorned standard model. However, we are interested
not in these fine details, but in the gross properties of the $X$ boson.
Therefore, we ignore mixing altogether and assume
\be
v_\U^2(q_\U-q_\Q)
+v_\D^2(q_\Q-q_\D)
=0\;.
\label{zeromixing}
\ee
For the same reason, we ignore mixing of the $B$ and the $X$
through the gauge boson kinetic energy terms, assuming that it is
negligible
throughout the range of energies of interest.

For (\ref{zeromixing}) to be satisfied,
$q_\U-q_\Q$ and $q_\D-q_\Q$ must have the same sign, which we take to be
positive by convention. Because $v_\U^2+v_\D^2\approx v^2$ (where $v
\simeq 246$~GeV is
the VEV of the standard model), we can write
\be
v_\U\approx v\,\sqrt{q_\D-q_\Q\over q_\U+q_\D-2q_\Q}\,,\quad\quad
v_\D\approx v\,\sqrt{q_\U-q_\Q\over q_\U+q_\D-2q_\Q}\,.
\label{vuandvd}
\ee
The contribution of the doublet VEVs to the mass of the $X$ boson is
\be
m_\X^{\rm min}\equiv g_\X\,v\,\sqrt{(q_\U-q_\Q)(q_\D-q_\Q)}\,.
\label{xmass}
\ee
Since there may also be $SU(2)$ singlet scalars contributing to the $X$
mass, (\ref{xmass})  should be regarded as a lower bound.

In the Higgs mechanism, one linear combination of the two charged fields,
$\xi_\U^\pm$ and $\xi_\D^\pm$ is transformed into the longitudinal
component of the $W^\pm$, while a similar linear combinations of the two
fields $\pi^0_\U$
and $\pi^0_\D$ becomes the
longitudinal component of the $Z$. If the $U(1)_\X$ breaking comes entirely
from the doublets, the other linear combination becomes
the
longitudinal component of the $X$.\footnote{We will discuss below what
happens if there is additional $U(1)_\X$ symmetry breaking.}
In unitary gauge, we may set
\be
\sum_{j\atop _{U,D}}\,v_j\xi_j^{+}
=\pi^0_\U=\pi^0_\D=0\;.
\ee
In our no-mixing, $v_\L=0$ approximation, we can take
\be
H_\U=
\pmatrix{
v_\D\,\xi^+/v\cr
(\xi^0_\U+v_\U)/\sqrt2\cr
}
\quad\quad
H_\D=
\pmatrix{
-v_\U\,\xi^+/v\cr
(\xi^0_\D+v_\D)/\sqrt2\cr
}
\label{doublets}
\ee
where $\xi^+$ is the surviving combination of $\xi_\U^+$ and
$\xi_\U^+$.
The $\xi^+$, $\xi^0_\U$ and $\xi^0_\D$ fields need not be mass
eigenstates 
(indeed,  we expect some mixing with the components of $H_\L$)
but for now we ignore mixing.

\section{\label{xdecays}$X$ Decays}

The couplings responsible for the decay of $X$ into $W$ or $Z$ plus a
scalar are obtained by
putting (\ref{doublets}) into (\ref{keterm}): 

\be\ba{c}
g_\X\,g_2\,{v_\U v_\D\over v}(q_\U+q_\D-2q_\Q)
\biggl(\xi^+\,X^\mu\,W^-_\mu
+\xi^-\,X^\mu\,W^+_\mu\biggr)\cra
+{g_\X\,g_2\over\cos\theta}
\biggl((q_\D-q_\Q)v_\D\,\xi^0_\U-(q_\U-q_\Q)v_\U\,\xi^0_\D\biggr)
X^\mu\,Z_\mu\;.\ea
\ee
Using (\ref{vuandvd}) and $m_\W=g_2\,v/2=m_\Z\cos\theta$, we find:
\be
2\,g_\X\,\sqrt{(q_\U-q_\Q)(q_\D-q_\Q)}\,
\biggl(m_\W\,\Bigr(\xi^+\,X^\mu\,W^-_\mu
+\xi^-\,X^\mu\,W^+_\mu\Bigr)
+m_\Z\,\xi^0\,X^\mu\,Z_\mu
\biggr)
\label{result}
\ee
where
\be
\xi^0\equiv\xi^0_\D\,\sqrt{q_\D-q_\Q\over q_\U+q_\D-2q_\Q}-
\xi^0_\U\,\sqrt{q_\U-q_\Q\over q_\U+q_\D-2q_\Q}\;.
\ee

Equation (\ref{result}), our central result, is valid provided that
all $SU(2)$ breaking is done by the VEVs of $H_\U$ and $H_\D$.  Our result
is unaffected by additional $U(1)_\X$ breaking due to the VEVs of
$SU(2)$ {\it singlet\/} fields. These would contribute to $m_\X$  and lead
to the survival of a linear combination of the various $\pi$ fields, but
one which
does not appear in (\ref{result}).

Note that $\xi^\pm$ and $\xi^0$ form a triplet under the custodial $SU(2)$
symmetry~\cite{custodial}. The orthogonal linear combination of the two
neutral states is a custodial $SU(2)$ singlet, which is the analog in this
model of the standard model Higgs boson. The custodial $SU(2)$ symmetry may
be broken by the mass mixing between the neutral states or by the masses and
mixing of all of the states with other spinless bosons in the model, but it
remains manifest in the couplings.

The dominant decay mode of $X$ is into quark-antiquark pairs, where the QCD
background may obscure the resonance. For this reason,
we are interested in the branching ratio for the decay of the $X$
into $W\xi$ and $Z\xi$ due to interaction
(\ref{result}). If the $\xi$s are sufficiently
light, these decays may offer clearer signatures of a leptophobic gauge
boson.

For each family, the rate
$\Gamma$ for $X$ to decay into a $Q={2\over3}$ quark-antiquark pair is:
\be\ba{l}
\Gamma(X\rightarrow U\ol U)\cra\approx
{1\over4\pi}\,g_\X^2\,(q_\Q^2+q_\U^2)\,(1-m_q^2/m_\X^2)\,p(m_\X,m_q,m_q)\cra
={1\over8\pi}\,g_\X^2\,(q_\Q^2+q_\U^2)\,(1-m_q^2/m_\X^2)\,\sqrt{m_\X^2-
4m_q^2}\;.\ea
\label{slg1}
\ee
where $p$ is the final particle momentum in the rest frame,
\be
p(m_\X,m_a,m_b)
={m_\X\over2}\,\sqrt{\left(1-{(m_a+m_b)^2\over m_\X^2}\right)
\left(1-{(m_a+m_b)^2\over m_\X^2}\right)}\,.
\ee
For each family, the rate
$\Gamma$ for $X$ to decay into a $Q=-{1\over3}$ quark-antiquark pair is:
\be\ba{l}
\Gamma(X\rightarrow D\ol D)\cra\approx
{1\over4\pi}\,g_\X^2\,(q_\Q^2+q_\D^2)\,(1-m_q^2/m_\X^2)\,p(m_\X,m_q,m_q)\cra
={1\over8\pi}\,g_\X^2\,(q_\Q^2+q_\D^2)\,(1-m_q^2/m_\X^2)\,\sqrt{m_\X^2-
4m_q^2}\;.\ea
\label{slg2}
\ee

One might expect the gauge boson decays to be
suppressed by powers of $m_\W/m_\X$ because of the explicit factors of
$m_\W$ and $m_\Z$ in (\ref{result}). However,
these factors are compensated
by the enhancement from the longitudinal $W$ and $Z$. Thus the partial
widths for $W\xi$, $Z\xi$ and $q\bar q$ decays are of the same order,
differing only by kinematic and counting factors.

For the $W^+$ decay, the square of the
invariant matrix element is
\be
{4\over3}
g_\X^2(q_\U-q_\Q)(q_\D-q_\Q)\,m_\W^2\,
\left(-g_{\mu\nu}+{{p_\W}^\mu {p_\W}^\nu\over m_\W^2}\right)
\left(-g_{\mu\nu}+{{p_\X}_{\!\mu} {p_\X}_{\!\nu}\over m_\X^2}\right)\;.
\ee
Using $(p_\X p_\W)=(m_\X^2+m_\W^2-m_\xi^2)/2$. we find
\be
{4\over3}g_\X^2\,(q_\U-q_\Q)(q_\D-q_\Q)\,m_\W^2\,
\left(2+{(m_\X^2+m_\W^2-m_\xi^2)^2\over4m_\X^2m_\W^2}\right)\;.
\ee
Thus the partial width
into $W^\pm\xi^\mp$ is
\be\ba{l}
\Gamma(X\rightarrow W^\pm\xi^\mp)\cra\approx
{1\over3\pi}g_\X^2\,(q_\U-q_\Q)(q_\D-q_\Q)\cra\cdot
\left(2+{(m_\X^2+m_\W^2-m_{\xi^\pm}^2)^2\over4m_\X^2m_\W^2}\right)
\,{m_\W^2\over m_\X^2}\,p(m_\X,m_\W,m_{\xi^\pm})\;.\ea
\label{gammaw}
\ee
For the decay into $Z\,\xi^0$, because of custodial symmetry,
the partial
width is given by half of this result, with
$m_\W\rightarrow
m_\Z$ and $m_{\xi^\pm}\rightarrow m_{\xi^0}$:
\be\ba{l}
\Gamma(X\rightarrow Z\xi^0)\cra\approx
{1\over6\pi}g_\X^2\,(q_\U-q_\Q)(q_\D-q_\Q)\cra\cdot
\left(2+{(m_\X^2+m_\Z^2-m_{\xi^0}^2)^2\over4m_\X^2m_\Z^2}\right)
\,{m_\Z^2\over m_\X^2}\,p(m_\X,m_\Z,m_{\xi^0})\;.\ea
\label{gammaz}
\ee

Branching ratios for the decay modes
$X\rightarrow W^\pm\xi^\mp$, $X\rightarrow Z\xi^0$ 
 are determined by equations
(\ref{slg1}), (\ref{slg2}), (\ref{gammaw}) and (\ref{gammaz}).
Figures \ref{fig121}, \ref{fig1212} and \ref{fig011} show these branching
ratios (and that of  $X\rightarrow t\bar t$)
as a function of $m_\X$
for two representative leptophobic models and two values for the $\xi$
masses. One model is that discussed in
reference \cite{hinchliffe}; in the other
we choose $q_\Q=0$, for which 
anomaly cancellation is more straightforward. The branching ratios for
these signature modes of decay of a leptophobic gauge boson are large
enough to be of experimental interest.

\section{\label{phenomenology}Phenomenology}

The production cross section for $X$ depends on $g_\X$, which otherwise
does not enter into our analysis except to determine
the minimum value of the $X$
mass, $m_\X^{\rm min}$. 
With $g_\X=0.15$, one of the values
discussed in reference \cite{hinchliffe}, this cross section is given
approximately in figure \ref{figsigma}. It is well below the published limit
from CDF for all values of $m_\X$. For this value of $g_\X$, $m_\X^{\rm
min}\approx 90$~GeV. 

The $\xi$'s produced in $X$ decays
decay primarily into heavy quarks, giving rise to the processes: 
{\renewcommand{\arraystretch}{.9}
\be
\begin{array}{r@{}l}
X\rightarrow W^- & \xi^+ \\
&\,\hookrightarrow
c\bar s
\ea
\ee
\be
\begin{array}{r@{}l}
X\rightarrow Z & \xi^0 \\
&\,\hookrightarrow
b\bar b
\end{array}
\ee
If, as expected,  the $\xi$'s mix slightly with the 
states in the $H_L$ doublet, 
there will also be decay modes in which $\tau$'s are
produced:
\be
\begin{array}{r@{}l}
X\rightarrow W^- & \xi^+ \\
&\,\hookrightarrow
\tau^+\nu_\tau
\ea\ee
\be
\begin{array}{r@{}l}
X\rightarrow Z & \xi^0 \\
&\,\hookrightarrow
\tau^+\tau^-
\end{array}
\ee
}
Note also that if the $X$ is above $t\bar t$ threshold, its decay into
$t\bar t$ could be a significant source of $t$s.

\section*{Acknowledgements}

We are grateful for interesting conversations with Melissa Franklin, Paolo
Giromini and Ken Lane. We are particularly grateful to Tom Baumann for help
with structure functions.

\newpage

{\figsize\begin{figure}[htb]
$$\beginpicture
\setcoordinatesystem units <\tdim,2000\tdim>
\setplotarea x from 200 to 500, y from 0 to .15
\put {$m_\X({\rm GeV})\rightarrow$} at 350 -.009
\put {$\displaystyle\uparrow\atop\displaystyle\%$} at 180 .075
\axis bottom visible ticks out quantity 4 /
\put {0} at 185 0
\put {5} at 185 .05
\put {10} at 185 .1
\put {15} at 185 .15
\axis left visible ticks out quantity 4 /
\put {200} at 200 -.007
\put {300} at 300 -.007
\put {400} at 400 -.007
\put {500} at 500 -.007
\stpltsmbl
\plot 200  .0982 202  .1006 204  .1026 206  .1046 208  .1062 210
.1078 212  .1092 214  .1104 216  .1116 218  .1126 220  .1134 222
.1142 224  .115 226  .1158 228  .1164 230  .1168 232  .1174 234
.1178 236  .1182 238  .1186 240  .119 242  .1194 244  .1196 246
.1198 248  .12 250  .1202 252  .1204 254  .1206 256  .1208 258
.1208 260  .121 262  .121 264  .1212 266  .1212 268  .1212 270
.1212 272  .1212 274  .1214 276  .1214 278  .1214 280  .1214 282
.1212 284  .1212 286  .1212 288  .1212 290  .1212 292  .1212 294
.121 296  .121 298  .121 300  .121 302  .1208 304  .1208 306
.1208 308  .1206 310  .1206 312  .1206 314  .1204 316  .1204 318
.1202 320  .1202 322  .1202 324  .12 326  .12 328  .1198 330
.1198 332  .1198 334  .1196 336  .1196 338  .1194 340  .1194 342
.1192 344  .1192 346  .119 348  .119 350  .1188 352  .1164 354
.1154 356  .1146 358  .1138 360  .1132 362  .1128 364  .1122 366
.1118 368  .1112 370  .1108 372  .1104 374  .11 376  .1096 378
.1094 380  .109 382  .1086 384  .1084 386  .108 388  .1078 390
.1074 392  .1072 394  .1068 396  .1066 398  .1064 400  .106 402
.1058 404  .1056 406  .1054 408  .105 410  .1048 412  .1046 414
.1044 416  .1042 418  .104 420  .1038 422  .1036 424  .1034 426
.1032 428  .103 430  .1028 432  .1026 434  .1024 436  .1022 438
.102 440  .1018 442  .1016 444  .1014 446  .1012 448  .1012 450
.101 452  .1008 454  .1006 456  .1004 458  .1004 460  .1002 462
.1 464  .0998 466  .0998 468  .0996 470  .0994 472  .0992 474
.0992 476  .099 478  .0988 480  .0988 482  .0986 484  .0984 486
.0984 488  .0982 490  .0982 492  .098 494  .0978 496  .0978 498
.0976 500  .0976 /
\setdashes
\plot 200 .0407 202 .0439 204 .0466  206 .0489 208 .0509 210 .0527
212 .0543 214 .0556 216 .0568  218 .0579 220 .0589 222 .0598  224
.0606 226 .0613 228 .0619  230 .0625 232 .0630 234 .0635  236
.0639 238 .0642 240 .0646  242 .0649 244 .0652 246 .0654  248
.0656 250 .0658 252 .0660  254 .0661 256 .0663 258 .0664  260
.0665 262 .0666 264 .0667  266 .0667 268 .0668 270 .0668  272
.0668 274 .0668 276 .0669  278 .0669 280 .0669 282 .0668  284
.0668 286 .0668 288 .0668  290 .0667 292 .0667 294 .0667  296
.0666 298 .0666 300 .0665  302 .0665 304 .0664 306 .0664  308
.0663 310 .0662 312 .0662  314 .0661 316 .0660 318 .0660  320
.0659 322 .0658 324 .0657  326 .0657 328 .0656 330 .0655  332
.0654 334 .0653 336 .0653  338 .0652 340 .0651 342 .0650  344
.0649 346 .0649 348 .0648  350 .0647 352 .0633 354 .0627  356
.0623 358 .0619 360 .0615  362 .0612 364 .0609 366 .0606  368
.0603 370 .0600 372 .0598  374 .0595 376 .0593 378 .0591  380
.0589 382 .0587 384 .0585  386 .0583 388 .0581 390 .0579  392
.0577 394 .0575 396 .0574  398 .0572 400 .0570 402 .0569  404
.0567 406 .0566 408 .0564  410 .0562 412 .0561 414 .0560  416
.0558 418 .0557 420 .0555  422 .0554 424 .0553 426 .0551  428
.0550 430 .0549 432 .0548  434 .0546 436 .0545 438 .0544  440
.0543 442 .0542 444 .0540  446 .0539 448 .0538 450 .0537  452
.0536 454 .0535 456 .0534  458 .0533 460 .0532 462 .0531  464
.0530 466 .0529 468 .0528  470 .0527 472 .0526 474 .0525  476
.0524 478 .0523 480 .0522  482 .0522 484 .0521 486 .0520  488
.0519 490 .0518 492 .0517  494 .0517 496 .0516 498 .0515  500
.0514 /
\setplotsymbol ({\large .})
\setdots <3\tdim>
\plot 350 .0005 352 .0200 354 .0280  356 .0341 358 .0392 360 .0436
362 .0475 364 .0511 366 .0544  368 .0575 370 .0604 372 .0631  374
.0657 376 .0682 378 .0705  380 .0727 382 .0749 384 .0770  386
.0789 388 .0809 390 .0827  392 .0845 394 .0862 396 .0879  398
.0896 400 .0912 402 .0927  404 .0942 406 .0957 408 .0971  410
.0985 412 .0998 414 .1012  416 .1025 418 .1037 420 .1050  422
.1062 424 .1073 426 .1085  428 .1096 430 .1107 432 .1118  434
.1129 436 .1139 438 .1150  440 .1160 442 .1170 444 .1179  446
.1189 448 .1198 450 .1207  452 .1216 454 .1225 456 .1234  458
.1242 460 .1251 462 .1259  464 .1267 466 .1275 468 .1283  470
.1290 472 .1298 474 .1306  476 .1313 478 .1320 480 .1327  482
.1334 484 .1341 486 .1348  488 .1355 490 .1361 492 .1368  494
.1374 496 .1381 498 .1387  500 .1393 /
\endpicture$$
\caption{\sf\label{fig121} Branching ratios in $X$ decay for
$(q_\Q,q_\U,q_\D)=(-
1,2,1)$ and $m_{\xi^\pm}=m_{\xi^0}=100$~GeV. The solid line is
$B(X\rightarrow W^\pm\,\xi^\mp)$. The dashed line is
$B(X\rightarrow Z\,\xi^0)$. The dotted line is
$B(X\rightarrow t\bar t)$.
}\end{figure}}

\newpage

{\figsize\begin{figure}[htb]
$$\beginpicture
\setcoordinatesystem units <\tdim,2000\tdim>
\setplotarea x from 200 to 500, y from 0 to .15
\put {$m_\X({\rm GeV})\rightarrow$} at 350 -.009
\put {$\displaystyle\uparrow\atop\displaystyle\%$} at 180 .075
\axis bottom visible ticks out quantity 4 /
\put {0} at 185 0
\put {5} at 185 .05
\put {10} at 185 .1
\put {15} at 185 .15
\axis left visible ticks out quantity 4 /
\put {200} at 200 -.007
\put {300} at 300 -.007
\put {400} at 400 -.007
\put {500} at 500 -.007
\stpltsmbl
\plot
280.2  .0005
 282  .01254
 284  .01773
 286  .02166
    288  .02490
 290  .02776
 292  .03006
 294  .03212
    296  .03406
 298  .03580
 300  .03742
 302  .03898
    304  .04038
 306  .04172
 308  .04298
 310  .04412
    312  .04538
 314  .04640
 316  .04752
 318  .04846
    320  .04940
 322  .05038
 324  .05126
 326  .05206
    328  .05290
 330  .05368
 332  .05444
 334  .05522
    336  .05590
 338  .05664
 340  .05734
 342  .05800
    344  .05858
 346  .05922
 348  .05984
 350  .06044
    352  .05966
 354  .05964
 356  .05976
 358  .05996
    360  .06018
 362  .06032
 364  .06060
 366  .06084
    368  .06100
 370  .06128
 372  .06150
 374  .06178
    376  .06202
 378  .06224
 380  .06246
 382  .06270
    384  .06292
 386  .06316
 388  .06332
 390  .06356
    392  .06380
 394  .06394
 396  .06416
 398  .06440
    400  .06458
 402  .06476
 404  .06496
 406  .06510
    408  .06538
 410  .06550
 412  .06564
 414  .06592
    416  .06608
 418  .06620
 420  .06640
 422  .06656
    424  .06676
 426  .06688
 428  .06706
 430  .06722
    432  .06736
 434  .06752
 436  .06764
 438  .06782
    440  .06796
 442  .06810
 444  .06822
 446  .06838
    448  .06852
 450  .06868
 452  .06876
 454  .06894
    456  .06900
 458  .06920
 460  .06930
 462  .06944
    464  .06954
 466  .06970
 468  .06978
 470  .06992
    472  .07006
 474  .07014
 476  .07028
 478  .07042
    480  .07044
 482  .07056
 484  .07074
 486  .07080
    488  .07088
 490  .07100
 492  .07114
 494  .07120
    496  .07134
 498  .07140
 500  .07144
/
\setdashes
\plot
 290.2  .0005
 292  .004792
 294  .008965
    296  .01165
 298  .01367
 300  .01539
 302  .01690
    304  .01817
 306  .01935
 308  .02045
 310  .02138
    312  .02235
 314  .02317
 316  .02400
 318  .02468
    320  .02539
 322  .02605
 324  .02668
 326  .02727
    328  .02783
 330  .02837
 332  .02887
 334  .02938
    336  .02984
 338  .03028
 340  .03071
 342  .03115
    344  .03153
 346  .03193
 348  .03230
 350  .03266
    352  .03226
 354  .03230
 356  .03239
 358  .03252
    360  .03264
 362  .03274
 364  .03289
 366  .03305
    368  .03315
 370  .03330
 372  .03342
 374  .03356
    376  .03371
 378  .03384
 380  .03394
 382  .03404
    384  .03420
 386  .03429
 388  .03439
 390  .03452
    392  .03461
 394  .03472
 396  .03481
 398  .03492
    400  .03503
 402  .03513
 404  .03519
 406  .03531
    408  .03540
 410  .03549
 412  .03555
 414  .03564
    416  .03574
 418  .03581
 420  .03590
 422  .03595
    424  .03604
 426  .03611
 428  .03619
 430  .03625
    432  .03630
 434  .03638
 436  .03646
 438  .03649
    440  .03658
 442  .03664
 444  .03669
 446  .03677
    448  .03681
 450  .03685
 452  .03691
 454  .03699
    456  .03702
 458  .03708
 460  .03712
 462  .03718
    464  .03724
 466  .03731
 468  .03733
 470  .03739
    472  .03743
 474  .03747
 476  .03752
 478  .03758
    480  .03758
 482  .03763
 484  .03771
 486  .03772
    488  .03775
 490  .03780
 492  .03786
 494  .03788
    496  .03794
 498  .03796
 500  .03795
/
\setplotsymbol ({\large .})
\setdots <3\tdim>
\plot
350  0
 352  .02218
 354  .03104
 356  .03770
 358  .04320
    360  .04799
 362  .05228
 364  .05612
 366  .05968
    368  .06298
 370  .06604
 372  .06896
 374  .07165
    376  .07425
 378  .07672
 380  .07908
 382  .08132
    384  .08346
 386  .08553
 388  .08754
 390  .08945
    392  .09131
 394  .09306
 396  .09483
 398  .09653
    400  .09811
 402  .09974
 404  .1012
 406  .1027
    408  .1042
 410  .1056
 412  .1070
 414  .1083
    416  .1096
 418  .1109
 420  .1121
 422  .1134
    424  .1145
 426  .1157
 428  .1168
 430  .1179
    432  .1190
 434  .1201
 436  .1212
 438  .1222
    440  .1232
 442  .1241
 444  .1251
 446  .1260
    448  .1269
 450  .1278
 452  .1287
 454  .1296
    456  .1305
 458  .1313
 460  .1321
 462  .1328
    464  .1336
 466  .1344
 468  .1352
 470  .1359
    472  .1366
 474  .1373
 476  .1381
 478  .1388
    480  .1395
 482  .1402
 484  .1409
 486  .1416
    488  .1421
 490  .1427
 492  .1434
 494  .1440
    496  .1448
 498  .1452
 500  .1458
/
\endpicture$$
\caption{\sf\label{fig1212} Branching ratios in $X$ decay for
$(q_\Q,q_\U,q_\D)=(-
1,2,1)$ and $m_{\xi^\pm}=m_{\xi^0}=200$~GeV. The solid line is 
$B(X\rightarrow W^\pm\,\xi^\mp)$. The dashed line is
$B(X\rightarrow Z\,\xi^0)$. The dotted line is
$B(X\rightarrow t\bar t)$.
}\end{figure}}

\newpage

{\figsize\begin{figure}[htb]
$$\beginpicture
\setcoordinatesystem units <\tdim,2000\tdim>
\setplotarea x from 200 to 500, y from 0 to .15
\put {$m_\X({\rm GeV})\rightarrow$} at 350 -.009
\put {$\displaystyle\uparrow\atop\displaystyle\%$} at 180 .075
\axis bottom visible ticks out quantity 4 /
\put {0} at 185 0
\put {5} at 185 .05
\put {10} at 185 .1
\put {15} at 185 .15
\axis left visible ticks out quantity 4 /
\put {200} at 200 -.007
\put {300} at 300 -.007
\put {400} at 400 -.007
\put {500} at 500 -.007
\stpltsmbl
\plot 200 .05608 202 .05756 204 .05890 206 .06018  208 .06120 210
.06222 212 .06308 214 .06386  216 .06466 218 .06522 220 .06586 222
.06640  224 .06694 226 .06732 228 .06776 230 .06810  232 .06838 234
.06878 236 .06906 238 .06930  240 .06950 242 .06970 244 .06980 246
.06998  248 .07018 250 .07028 252 .07040 254 .07050  256 .07060 258
.07072 260 .07078 262 .07086  264 .07082 266 .07092 268 .07092 270
.07096  272 .07094 274 .07102 276 .07100 278 .07100  280 .07094 282
.07096 284 .07102 286 .07096  288 .07092 290 .07092 292 .07090 294
.07086  296 .07076 298 .07078 300 .07070 302 .07064  304 .07062 306
.07066 308 .07060 310 .07056  312 .07048 314 .07046 316 .07038 318
.07032  320 .07028 322 .07024 324 .07014 326 .07014  328 .07010 330
.06996 332 .06992 334 .06988  336 .06980 338 .06970 340 .06970 342
.06962  344 .06954 346 .06950 348 .06948 350 .06938  352 .06838 354
.06792 356 .06752 358 .06728  360 .06696 362 .06670 364 .06648 366
.06630  368 .06602 370 .06586 372 .06570 374 .06548  376 .06528 378
.06520 380 .06494 382 .06478  384 .06464 386 .06448 388 .06432 390
.06420  392 .06410 394 .06388 396 .06376 398 .06366  400 .06352 402
.06342 404 .06328 406 .06314  408 .06308 410 .06296 412 .06276 414
.06266  416 .06262 418 .06246 420 .06238 422 .06230  424 .06212 426
.06204 428 .06194 430 .06188  432 .06174 434 .06168 436 .06158 438
.06142  440 .06140 442 .06130 444 .06118 446 .06112  448 .06098 450
.06094 452 .06086 454 .06078  456 .06064 458 .06062 460 .06054 462
.06046  464 .06038 466 .06038 468 .06022 470 .06016  472 .06016 474
.06004 476 .05994 478 .05990  480 .05980 482 .05974 484 .05964 486
.05964  488 .05954 490 .05944 492 .05948 494 .05928  496 .05928 498
.05920 500 .05916 /
\setdashes
\plot 200 .02322 202 .02510 204 .02674 206 .02813  208 .02935 210
.03041 212 .03132 214 .03218  216 .03297 218 .03359 220 .03419 222
.03473  224 .03524 226 .03566 228 .03602 230 .03639  232 .03672 234
.03700 236 .03728 238 .03750  240 .03772 242 .03791 244 .03806 246
.03819  248 .03835 250 .03846 252 .03855 254 .03867  256 .03876 258
.03884 260 .03890 262 .03895  264 .03900 266 .03904 268 .03907 270
.03908  272 .03908 274 .03914 276 .03914 278 .03911  280 .03909 282
.03910 284 .03914 286 .03910  288 .03909 290 .03904 292 .03903 294
.03903  296 .03898 298 .03896 300 .03892 302 .03887  304 .03882 306
.03881 308 .03879 310 .03874  312 .03872 314 .03863 316 .03864 318
.03857  320 .03852 322 .03847 324 .03839 326 .03837  328 .03832 330
.03827 332 .03819 334 .03820  336 .03817 338 .03805 340 .03801 342
.03798  344 .03792 346 .03787 348 .03782 350 .03776  352 .03719 354
.03693 356 .03670 358 .03652  360 .03633 362 .03618 364 .03602 366
.03590  368 .03577 370 .03568 372 .03557 374 .03542  376 .03536 378
.03521 380 .03506 382 .03499  384 .03494 386 .03481 388 .03471 390
.03460  392 .03454 394 .03442 396 .03431 398 .03427  400 .03416 402
.03407 404 .03398 406 .03390  408 .03388 410 .03376 412 .03365 414
.03360  416 .03358 418 .03345 420 .03337 422 .03335  424 .03324 426
.03317 428 .03310 430 .03304  432 .03295 434 .03294 436 .03288 438
.03278  440 .03274 442 .03268 444 .03258 446 .03255  448 .03247 450
.03244 452 .03237 454 .03231  456 .03226 458 .03222 460 .03214 462
.03211  464 .03205 466 .03200 468 .03192 470 .03188  472 .03185 474
.03180 476 .03174 478 .03169  480 .03163 482 .03157 484 .03156 486
.03151  488 .03143 490 .03140 492 .03136 494 .03130  496 .03125 498
.03122 500 .03120 /
\setplotsymbol ({\large .})
\setdots <3\tdim>
\plot350 0 352 .01411 354 .01983 356 .02417 358 .02778  360 .03095
362 .03378 364 .03635 366 .03877  368 .04098 370 .04309 372 .04506
374 .04688  376 .04869 378 .05043 380 .05201 382 .05359  384
.05513 386 .05654 388 .05797 390 .05933  392 .06069 394 .06189 396
.06311 398 .06438  400 .06550 402 .06665 404 .06776 406 .06879  408
.06985 410 .07093 412 .07194 414 .07287  416 .07382 418 .07477 420
.07570 422 .07658  424 .07749 426 .07835 428 .07921 430 .08000  432
.08080 434 .08156 436 .08236 438 .08313  440 .08386 442 .08458 444
.08529 446 .08602  448 .08664 450 .08744 452 .08808 454 .08872  456
.08942 458 .09008 460 .09063 462 .09127  464 .09188 466 .09248 468
.09300 470 .09362  472 .09420 474 .09475 476 .09529 478 .09585  480
.09638 482 .09690 484 .09749 486 .09793  488 .09844 490 .09888 492
.09944 494 .09993  496 .1004 498 .1008 500 .1013 /
\endpicture$$
\caption{\sf\label{fig011} Branching ratios in $X$ decay for
$(q_\Q,q_\U,q_\D)=(
0,1,1)$ and $m_{\xi^\pm}=m_{\xi^0}=100$~GeV. The solid line is
$B(X\rightarrow W^\pm\,\xi^\mp)$. The dashed line is
$B(X\rightarrow Z\,\xi^0)$. The dotted line is
$B(X\rightarrow t\bar t)$.
}\end{figure}}

\newpage

{\figsize\begin{figure}[htb]
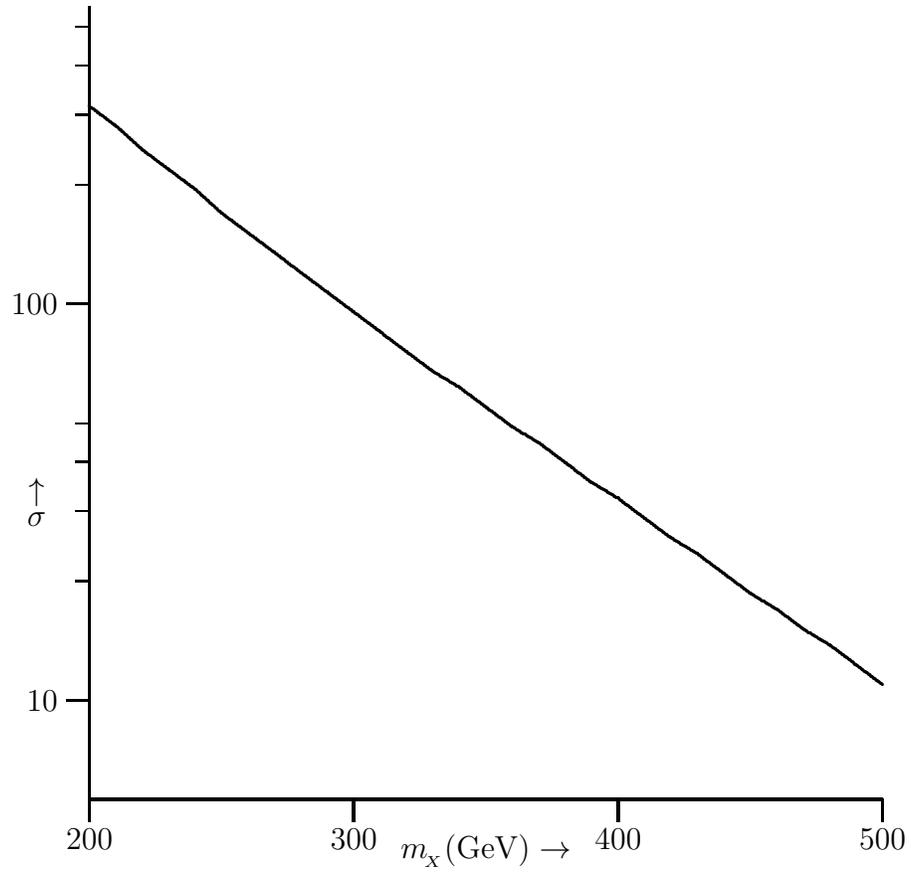

$$\beginpicture
\setcoordinatesystem units <\tdim,150\tdim>
\stpltsmbl
\setplotarea x from 200 to 500, y from .75 to 2.75
\linethickness=.013truein
\axis bottom ticks long out quantity 4 /
\axis left ticks long out logged unlabeled at 10 100 / /
\put {$10$} [r] at 188 1
\put {$100$} [r] at 188 2
\put {$\displaystyle\uparrow\atop\displaystyle\sigma$} at 180 1.5
\put {200} at 200 .65
\put {300} at 300 .65
\put {400} at 400 .65
\put {500} at 500 .65
\put {$m_\X({\rm GeV})\rightarrow$} at 350 .62
\linethickness=.006truein
\axis left ticks short out logged unlabeled at 20 200 / /
\axis left ticks short out logged unlabeled at 30 300 / /
\axis left ticks short out logged unlabeled at 40 400 / /
\axis left ticks short out logged unlabeled at 50 500 / /
\plot 200  2.5 210  2.45 220  2.39 230  2.34 240  2.29 250  2.23
260  2.18 270  2.13 280  2.08 290  2.03 300  1.98 310  1.93 320
1.88 330  1.83 340  1.79 350  1.74 360  1.69 370  1.65 380  1.6
390  1.55 400  1.51 410  1.46 420  1.41 430  1.37 440  1.32 450
1.27 460  1.23 470  1.18 480  1.14 490  1.09 500  1.04 /
\endpicture$$
\caption{\sf\label{figsigma}An estimate of the $X$ production cross section
in
picobarns at center of mass energy 1800 GeV for $(q_\Q,q_\U,q_\D)=(-1,2,1)$
and $g_\X=0.15$.}\end{figure}}

\end{document}